\documentclass[12pt,preprint]{aastex}

\newcommand{\es}{1ES1959$+$65}

\shorttitle{X--ray Variability of \es}
\shortauthors{Giebels et al.}

\begin{document}

\title{Observation of X--ray variability \\
in the BL Lac object \es}
\author{  
Berrie Giebels\altaffilmark{1},  
Elliott D. Bloom,  
Warren Focke,  
Gary Godfrey, Greg Madejski, 
Kaice T. Reilly,  
Pablo M. Saz Parkinson, Ganya Shabad 
}  
\affil{Stanford Linear Accelerator Center, Stanford University,   
Stanford, CA 94309}  
\author{    
Reba M. Bandyopadhyay\altaffilmark{2},
Gilbert G. Fritz,  
Paul Hertz\altaffilmark{3},
Michael P. Kowalski,\\
Michael N. Lovellette,
Paul S. Ray,
Michael T. Wolff,
Kent S. Wood, 
Daryl J. Yentis
}  
\affil{E. O. Hulburt Center for Space Research, Naval Research  
Laboratory, Washington, DC 20375}  
\and  
\author{Jeffrey D. Scargle}  
\affil{Space Science Division, NASA/Ames Research Center,   
Moffett Field, CA 94305-1000}  
\altaffiltext{1}{Berrie.Giebels@slac.stanford.edu, current address: Laboratoire Leprince-Ringuet, Ecole Polytechnique, Palaiseau F-91128}    
\altaffiltext{2}{NRL/NRC Research Associate}  
\altaffiltext{3}{Current address: NASA Headquarters, 300 E Street, SW, Washington,  
DC\ 
 20546-0001}  

\begin{abstract}
This paper reports X--ray spectral observations of a relatively nearby 
($z = 0.048$) BL Lacertae (BL Lac) object \es, which is a potential TeV 
emitter.  The observations include 31 short 
pointings made by the Unconventional Stellar Aspect (USA) Experiment on
board the {\em Advanced Research and Global Observation Satellite}
(ARGOS), and 17 pointings by the PCA on board the {\em Rossi X--ray Timing
Explorer} (RXTE). Most of these  observations were spaced by less than
1 day. \es\  was detected by the ARGOS USA detector in the range 1-16 keV, 
and by the PCA in the 2-16 keV range but at different times. During the 
closely spaced RXTE observations beginning on 2000 July 28 , an ending of one flare and a start 
of another are visible, associated with spectral changes, where the 
photon index $\Gamma$ ranges between $\sim$ 1.4 and 1.7, and the 
spectrum is harder when the source is brighter.  This implies 
that \es\ is an XBL-type blazar, with the X--ray emission likely 
to originate via the synchrotron process.  The USA observations  
reveal another flare that peaked on 2000 November 14 and doubled the 
flux within a few days, again associated with spectral changes  
of the same form.  The spectral variability correlated with the 
flux and timing characteristics of this object that are 
similar to those of other nearby BL Lacs, and suggest relativistic 
beaming with a Doppler factor $\delta \geq 1.6$ and magnetic fields 
of the order of a few mG. We also suggest that the steady component 
of the X--ray emission -- present in this object as well as in other 
XBLs -- may be due to the large-scale relativistic jet (such as 
measured by {\em Chandra} in many radio-loud AGN), but pointing 
very closely to our line of sight.  

\end{abstract}

\section{Introduction}

Over a dozen BL Lacs have been detected at GeV energies \citep{mukh},
but only a few nearby BL Lacs have been identified at TeV energies so
far. Mkn 421 ($z=0.031$) \citep{punch} and Mkn501 ($z=0.034$)
\citep{quinn} are now strongly confirmed sources, seen by more than 
one ground-based atmospheric \v{C}erenkov telescope (ACT) at or
above the $5 \sigma$ level. Two more, 1ES 2344+514 (z=$0.044$) \citep{cata} 
and PKS 2155-304 ($z=0.116$) \citep{chad}, have been detected only once 
and are less conclusive. This strongly suggests that low-redshift 
X--ray selected BL Lac objects (XBLs) such as these may be the only 
extragalactic $\gamma$-ray sources observable at TeV energies. 
This is because on the one hand more distant objects would have their
TeV emission   
strongly absorbed by the extragalactic background light (EBL), 
and on the other bright X--ray emission, presumably originating from synchrotron
radiation and thus revealing the distribution of radiating particles, 
implies even higher intrinsic GeV-TeV emission (see, e.g. \citet{tave}). 

The BL Lac object \es\ ($z=0.048$), is an XBL also present in the third
EGRET catalog with an average measured flux of $1.8\times10^{-7}$ photons $
\textrm{cm}^{-2} \textrm{s}^{-1}$ for $E>100$ MeV \citep{hart}. It is
thus a natural source for TeV emission, and \citet{steck}, using 
simple scaling arguments, have predicted for it the third highest 
flux above 0.3 and 1 TeV, after Mkn 421 and Mkn 501. More recently, 
\citet{costaghi} have also pointed it out as a candidate TeV emitter. 
Quoting \citet{weekes}, the Utah Seven Telescope Array has reported the detection of \es\ 
based on 57 hours of observation in 1998 \citep{kaj}, with an energy 
threshold of 600 GeV. The flux level was not reported but the total 
signal was at the 3.9 {$\sigma$} level. This is not normally considered 
sufficient to claim the detection of a new source; however, within 
this database there were two epochs which were selected 
\textit{a posteriori} which gave signals above the canonical 
{$5\sigma$} level. This source has not yet been confirmed
by any other group; it was observed by the Whipple group but no flux
was detected \citep{catan}.  Of particular interest in this source is
also the fact that its distance is of the same order as the two other
confirmed TeV sources, making it a good candidate to probe the
EBL that can in turn probe cosmological problems such as the formation 
of galaxies (see e.g. \citet{prim,guy}).  
\es\ is also part of a 200 mJy radio-selected sample at 5 GHz
\citep{marcha,bondi} and it  
was seen in the Einstein Slew Survey \citep{perlm}.  Multiple
photometric optical 
values were found in the literature, showing a great variation of 
the source brightness in the optical band from $V=16$ to $V=12.8$.  
A complete study of the optical band can be found in \citet{villa} 
where variability on short timescales (a few days) was reported.

The capabilities of the USA and the PCA instruments to monitor the 
X--ray emission are particularly well suited for the detailed
study of the X--ray energy spectrum of \es, and its temporal 
evolution. Here, we present 31 USA and 17 PCA observations of \es.  
The PCA data are two sets of intraday observations, obtained from 
unpublished RXTE archival data, that span a few days in which the 
end of a flare and beginning of another flare are detected.  
The USA observations are short daily monitoring observations 
that span 7 weeks, ending with a strong flare, where the 
X--ray flux tripled in 7 days (the USA data were taken in the 
context of a multi-wavelength campaign that was not successful 
in the TeV range due to bad weather at the ACT site).  In Section 2, 
we report the details of the observations;  in Section 3, we report 
the main features of the data;  in Section 4, 
we discuss the implications of the X--ray spectral variability 
and the quiescent X--ray emission detected in \es;  we present 
our conclusions in Section 5.  

\section{Observations and Data Analysis}

\subsection{USA Data}

The USA Experiment observed \es\ about once per day between
2000 September 21 (MJD 51808) and November 11 (MJD 51859). 
(For a detailed description of the USA experiment see \citet{psr},
\citet{wood} and \citet{sha}.)   The USA detector was used in the ``ping'' mode 
where within one observation the source is pointed at 2 or 3 times 
for $T_{ON}$ seconds and then a background is taken for $T_{OFF}$ 
seconds, where typically, $T_{ON}$ = $T_{OFF}$ = 60 seconds and an 
observation lasted for $\sim$300 s (for more details on the ``ping'' 
observation mode see \citet{ber}). This type of observation
has the advantage of not relying on a background model but rather on
a direct measurement of the background, but results in some loss 
of signal within an observation period.  USA data were extracted from 
FITS formatted files using CFITSIO.  Care was taken to use an 
$OFF$ position as devoid as possible of bright X--ray sources, 
and at least 2$^\circ$ away from the source.  Observations were made 
in low background regions of the orbit, where the counting rate was 
approximately 30 cts/s or 1\% of the Crab level in the same conditions. 
The segments of the observation where the background is too high, 
especially at the beginning and the ending of each observation, 
are rejected.  For the present investigation, we use only observations 
where at least 2 pairs of ON - OFF pointings are available for each 
observation.  Within those restrictions, 31 observations were used 
from which $\sim 16$~ks of data were selected.  

\begin{figure} 
\begin{center}
\includegraphics[angle=0,scale=0.6]{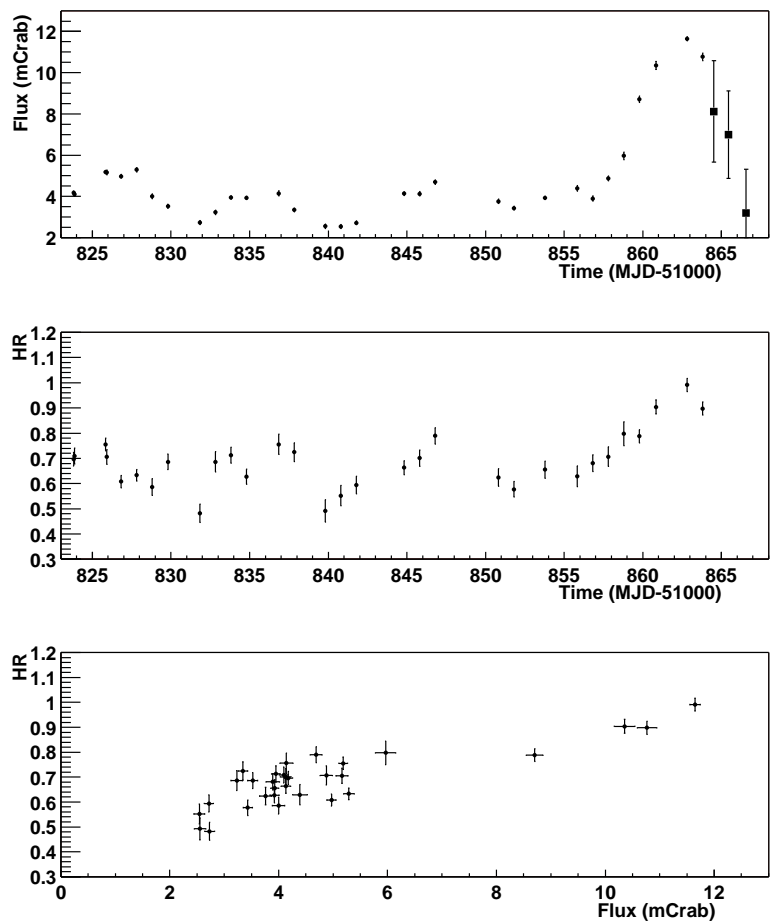}
\end{center}
\caption{Time evolution of \es\ and hardness ratio seen by
USA. Top panel: filled circles are the flux (in units of milliCrab),
the filled squares are the flux seen by the RXTE/ASM and are displayed 
to complete the observation of the largest flare when the USA
observations ceased.The normalization to milliCrabs for the ASM was
done using 1 Crab = 75 ASM counts/s. The errors shown here are only
statistical. The conversion into milliCrabs has an additional 6\%
systematic error due to a 0.1$^\circ$ systematic error in the pointing 
of the instrument.
Middle panel: 
Hardness ratio (HR) as a function of time.
Bottom panel: HR as a function of the observed flux.}
\label{plotone}
\end{figure}

We present the X--ray lightcurve for \es\ in Figure \ref{plotone};  
in that Figure, the standard deviation of the average in the
background was added in quadrature to the error on the count rate for
each observation.  The 
data were then corrected for obscuration by the instrument support
structure when necessary, and also for the collimator response.  
Every point in the lightcurve is a single observation. The lightcurve 
(normalized to the USA Crab rate in the energy range defined below, 
or 3500 cts/s) for the total range is shown in the top panel of 
Figure \ref{plotone}. 

The data were taken in the spectral mode, where the instrument integrates 
a spectrum covering an energy range of approximately 1--17~keV in 48 
pulse height analyzer (PHA) channels every 10 ms. In this work, we
make no use of 
the lowest (0) and the highest (47) PHA channels;  the PHA channels 
1--46 ($\sim 1-16$~keV) are referred to as the total USA range. 
The spectral characteristics of the time series were studied by
dividing the USA data into two energy bands, the PHA channels 1--10
(soft band) and 11--46 (hard band), corresponding to approximately 
1--3 keV and 3--16 keV, respectively. A hardness ratio, shown in the 
second from top panel in Figure \ref{plotone}, is the ratio of the 
counting rate in the hard band over the soft band.  The dates are 
given in Modified Julian Date (MJD) $-$ 51000.  

\subsection{RXTE/PCA Data}

Unpublished PCA data of \es\ were obtained from the RXTE data archive.  
The RXTE/PCA observed \es\ 12 times from 2000 July 28 through August
2, and 5 times between 2000 September 1--6, with exposures of
$\sim$900 s in the July-August observations and a few ksec each in
September. The STANDARD2 data were extracted using the HEASARC ftools and 
filtered using the RXTE GOF-recommended criteria (layer 1 only, for better 
signal to noise, PHA channels 0 to 27 (lightcurves only) or approximately
1-10 keV, Earth elevation angle greater than 10 degrees, pointing
offset less than 0.02 degrees, time since the peak of the last SAA passage 
greater than 30 minutes and electron contamination less than 0.1).  
The lightcurve and spectral data are from unit 2 (PCU 2) only since a faint background model is 
not yet available for unit 0 during gain epoch 5. The background 
models of Epoch 4 were used. Lightcurves were extracted using the 
ftools \texttt{saextrct} (through the \texttt{rex} script) and 
\texttt{lcurve}. The variable PCA background was modeled with 
\texttt{pcabackest}, which uses observations of X--ray blank, high 
latitude areas of the sky \citep{jaho}. Spectral fits were done 
using XSPEC v. 11.0.1 and response matrices generated by the ftool 
\texttt{pcarmf}. 

The source was not detected with a better significance than {$\sim
2\sigma$} with the RXTE/HEXTE instruments. A likely explanation is the
steep spectrum, derived from the PCA data, that falls below the
HEXTE sensitivity. 

\subsection{Archival Data}

 An observation with the \textit{BeppoSAX} instrument in 1997  
\citep{beck,beck2} resulted in a measured flux of $1.3\times10^{-11}$ 
erg cm$^{-2}$ s$^{-1}$ in the 2--10 keV band and a spectral index of
{$\alpha=1.64$}. This flux is an order of magnitude fainter
than the brightest flux measured here by the PCA. However, the results
published by \citep{beck} should be treated with caution, as there is
an apparent error in the value of the Galactic column  
adopted by them (they adopt $10^{20}$ rather than $1.027 \times 
10^{21}$ atoms cm${-2}$ adopted by us on the basis of the COLDEN 
program available as a part of the Chandra Proposal Planning Toolkit, 
based on relatively reliable 21 cm data).  The analysis of the PSPC 
All-Sky Survey data by those authors implies that assuming a 
simple power law, the fitted absorption is $1.6 \times 10^{21}$ atoms 
cm${-2}$, somewhat larger than the Galactic value.  However, this might be 
because the intrinsic X-ray spectrum steepens somewhat towards higher 
energies, as is often the case for other XBL-type blazars, and an assumption 
that the observed spectrum is a simple absorbed power law overestimates 
the fitted absorption. Note that the higher column density makes no difference
in the results from the USA or RXTE data. The \textit{Einstein} Slew Survey Sample of BL Lac Objects \citep{perlm} quotes a flux
of 3.65 $\mu$Jy at 2 keV, which is $\sim$40\% brighter than the
\textit{BeppoSAX} measurement.  

We also extracted the \textit{ROSAT} HRI archival data for this source, 
and used the \textit{ROSAT} all-sky data.  
\textit{ROSAT} HRI observed it on 1996 April 1; the observation lasted 
for a total of half a day, yielding about 2800 sec of good data.    
The data were reduced in a standard manner, revealing that the net 
source counting rate was $\sim$ 1.57 ct s$^{-1}$, with no indication 
of variability, but this is not too surprising given the short 
observation length.  The conversion of the HRI count rate to flux 
is dependent on the source spectrum, which has to be assumed as there 
is essentially no spectral information in the HRI data. Since we do not know the soft X-ray spectrum at the epoch of the 
HRI observation, we assume the energy index $\alpha = 1.5$, but 
two different values of $N_H$, the Galactic value of $10^{21}$ 
cm$^{-2}$ and the fitted ROSAT value of $1.6 \times 10^{21}$ 
cm$^{-2}$.  To obtain the conversion from the HRI counting rate to 
the observed flux, we used the PIMMS tool provided by HEASARC (and checked the results using XSPEC with the HRI effective 
area curve).  Assuming the Galactic column of $10^{21}$ 
cm$^{-2}$, we obtain the 0.1 - 2.4 keV flux of $5.5 \times 10^{-11}$ 
erg cm$^{-2}$ s$^{-1}$ and 1 - 2 keV flux of $2.2 \times 10^{-11}$ 
erg cm$^{-2}$ s$^{-1}$.  Assuming the fitted ROSAT value of 
$1.6 \times 10^{21}$ cm$^{-2}$, we infer the 0.1 - 2.4 keV 
flux of $5.4 \times 10^{-11}$ erg cm$^{-2}$ s$^{-1}$ and 1 - 2 keV flux 
of $2.5 \times 10^{-11}$ erg cm$^{-2}$ s$^{-1}$.  In any case, 
this corresponds to (roughly) 2 mCrab, which is lower than 
the 13 mCrab level from the \textit{Rosat all-sky Survey 
Bright Source Catalog} (1RXS-B).  This simply implies that during
the \textit{ROSAT} survey, \es\ was significantly brighter than during 
the \textit{Einstein}, \textit{ROSAT} HRI or \textit{BeppoSAX} 
observations, and the episodes of high flux as seen by the USA 
or RXTE observations described hereafter are not unique.

\section{Observational Results}

\subsection{Flares in the USA and RXTE Data}

The USA and RXTE observations of \es\ conducted from July through
November 2000 show that the source was bright and variable in the
X--ray band, with the X--ray spectrum significantly harder than observed 
during  the periods of lower brightness. Specifically, these data 
show that during the last quarter of 2000, \es\ underwent an X--ray flare
reaching the 12 mCrab level in the 1-16 keV band on November 14 
(MJD 51863). Variability of a factor $\sim$6 was detected within 20 days, 
and a factor $\sim$ 3 within 7 days (Figure~\ref{plotone}). By comparison,
the peak flux detected by USA on Mkn 421 in 2000 reached approximately
40 mCrab at maximum. Visual inspection of the USA (as well as the RXTE) 
lightcurves, and in particular of the largest flare, indicates that 
the source does not appear to vary significantly on timescales shorter 
than a day; thus the variability is not undersampled.
These observations show that \es\ was in a variable state for at 
least 4 months. To complete the coverage of this flare, 3 data points
from the RXTE/ASM were normalized to the Crab and added to the lightcurve 
in Figure~\ref{plotone};  this shows the flare continuing to decrease. 
The varying spectral index and differences in the energy response of both
instruments complicate the comparison. Nonetheless, the full-width at 
half-maximum region of the flare with these additional points spans 
5 $\pm$ 1 day, and the doubling time is 2.5 days.

\begin{figure} 
\begin{center}
\includegraphics[angle=0,scale=0.6]{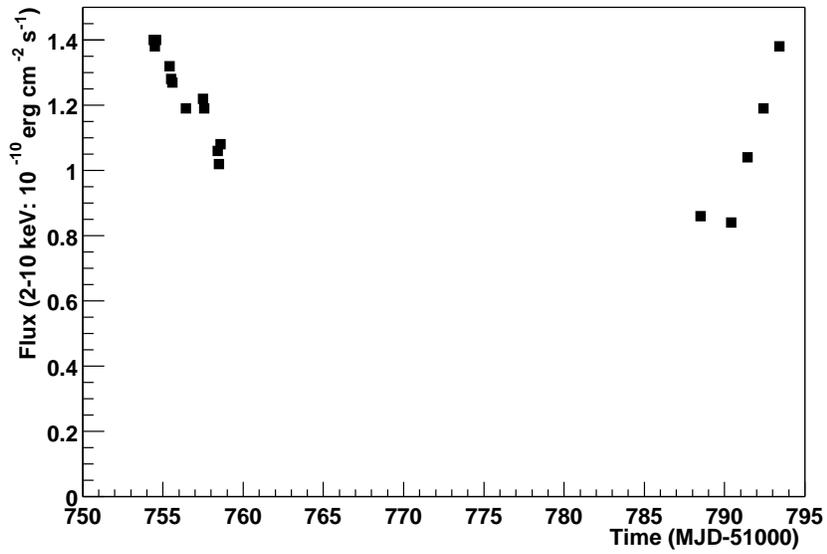}
\end{center}
\caption{  
Time history of \es\ emission obtained from the PCA. The first part of
the observations go from 
July 28 to Aug 2 (2000), and the second part from Sep 1 to Sep 4. For
comparison with Fig. 1, a flux level of 1 mCrab is approximately
$1.7\times10^{-11}$ erg cm$^{-2}$
s$^{-1}$ in the 2--10 keV band.}
\label{plottwo}
\end{figure}

\begin{figure} 
\begin{center}
\includegraphics[angle=270,scale=0.5]{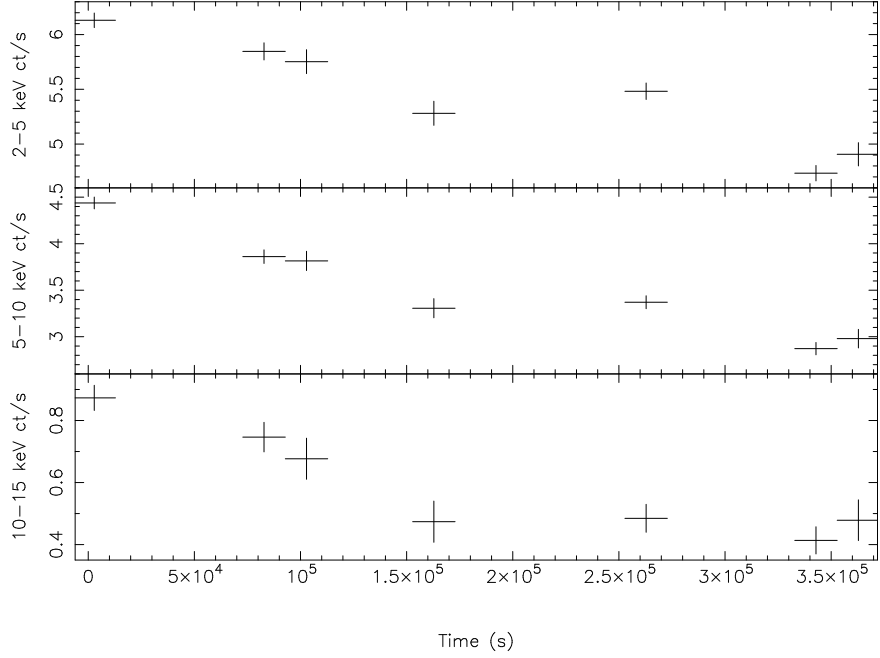}
\end{center}
\caption{  
RXTE PCA count rates for \es\ during the first
(decreasing) part of the observations. Count rates are given in the
2--5 keV (top), 5--10 keV (second from top) and 10--15 keV bands (third
from top). Given the observed spectra, the weighted
mean energy of the photons in the 2--5 keV band is $\sim$3 keV, in the 
5--10 keV band it is $\sim$7 keV, and in the 10--15 keV band it is
$\sim$12 keV. These values are spectral index dependent, but change
only a few percent for the spectra observed. The timescale is the same 
as in figure \ref{plottwo}. In the 2--5 keV and the 5--10 keV bands
the drop is 24\% and 35\% respectively in $3.5\times10^5$ s. In the 10--15 keV range a plateau is
reached after $1.7\times10^5$ s after a 46\% drop in flux, thus the
halving time is taken here to be $3.7\times10^5$ s.} 
\label{plotthree}
\end{figure}

The PCA archival data obtained 2 months prior to the USA observations show 
65\% flux changes in 3.5 days;  the highest observed value was  
$F$(2-10 keV) = $1.4\times10^{-10}$ erg cm$^{-2}$ s$^{-1}$, but the
peak value is unknown, since the observed maxima are at the endpoints
of the observed period, when the source was falling or rising as shown
in Figure~\ref{plottwo}. The USA 
lightcurve shows that the PCA did not cover the typical variation
period which appears to be $>$ 4 days. The same figure shows a decreasing 
flux extending over 3.5 days, and after 30 days observations resume 
for 4 days where a steady increase of flux is seen. The flux did not 
change more than a few percent on timescales shorter than a day in 
the PCA data. In the three panels in Figure \ref{plotthree} it is
apparent that variations are larger in the harder bands in the
decreasing part of the PCA observations.    

\subsection{Flux-Spectrum Correlations}
The PCA data were used to perform spectral fits as a function of
the flux in the 2-10 keV energy range. The data were fit to a single power-law function with 
index {$\alpha$}, such that the photon flux {$N(E) = N_0 E^{-(\alpha+1)}$} 
and the absorbing column $N_H$=10.27 $\times 10^{20}$ cm$^{-2}$.

 The absorbed power law
model provides an adequate fit for all   
RXTE PCA pointings. Spectral indices were obtained 
for every observation, and in some cases, intraday observations where 
the estimated indices and fluxes were similar were added together to 
improve the significance on a daily timescale.  The spectral fits are
shown in Table 2 along with the month, day and fraction of the day of
the beginning of each observation. 

\begin{figure} 
\begin{center}
\includegraphics[angle=0,scale=0.6]{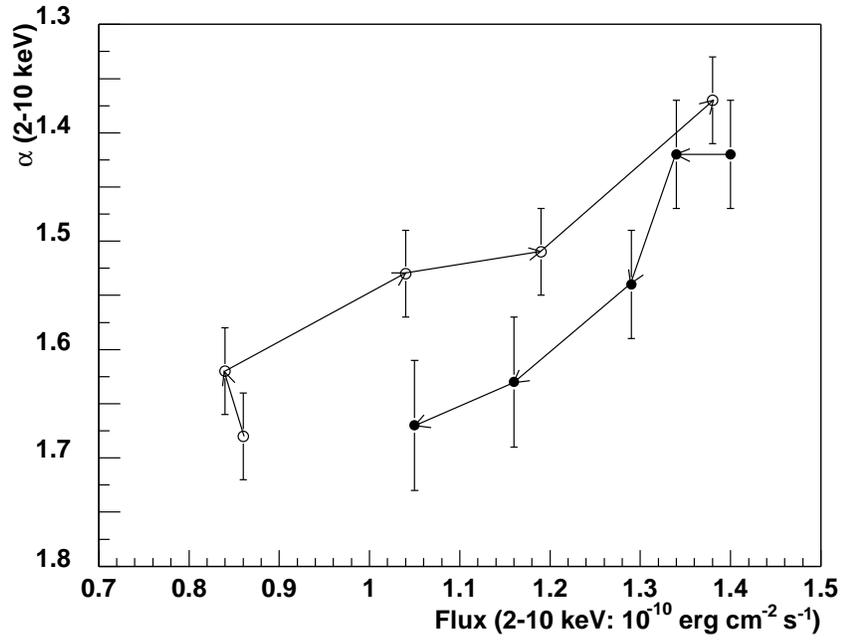}
\end{center}
\caption{
Evolution of the X--ray spectral index $\alpha$ ($F_\nu \propto
\nu^{-\alpha}$) of \es\ as a function of the X--ray flux in the PCA
data. The model is  
a power law with fixed absorption column density $N_H$=10.27$^{20}$
cm$^{-2}$. Filled circles: declining phase. Open circles: rising
phase. Every point is a daily average from the data in Table 1.} 
\label{plotfour}
\end{figure}

The X--ray spectrum follows a ``loop'' in the spectral index-flux
plane, as seen in Figure \ref{plotfour}, and it is not surprising that 
the X--ray spectrum shows significant evolution during the flare 
given that there is a more rapid rise and drop in the hard X--ray band 
(see \S \ref{par:disc} for discussion). The steepest spectrum in the
PCA data was observed at the first observations in the declining phase 
July 28 ({$\alpha = 1.68$}), and the hardest spectrum was seen on 2
dates separated by a month and at a similar flux ({$\alpha = 1.37$})
which is a hint that the same physical mechanism is generating these 
flux variations.  In the case of the USA observation, poorer photon 
statistics and an incomplete energy calibration limited the spectral 
study to a hardness ratio (HR) estimation plotted on the same Figure
\ref{plotone}. The USA fluxes and HR are shown in table 1. During the strongest flare, that started around
(MJD-51000)=857, a $20 \%$ variation in the HR is observed.

\section{Discussion \label{par:disc}}

\subsection{Doppler Boosting of the Flux of \es}

The electromagnetic emission in blazars is very likely to be
Doppler-boosted (or beamed) towards the observer.  In the radio regime, 
the evidence comes from superluminal expansions observed with
VLBI.  Similar superluminal expansions have now been seen in the optical 
band with the \textit{HST} in nearby galaxies such as M87. Relativistic 
beaming is also required in order to avoid absorption of GeV photons 
by X--ray photons via the e$^{+}$/e$^{-}$ pair-production process. 
It is thus possible to use the X--ray variability data as well as the fact 
that \es\ is an EGRET-detected BL Lac object, to establish a limit 
for the Doppler factor {$\delta$}, with {$\delta$} defined in 
the standard way as {$[\Gamma(1-\beta\cos \theta)]^{-1}$}, where 
{$\Gamma$} is the bulk Lorentz factor of the plasma in
the jet, {$\beta=v/c$} and {$\theta$} is the angle to the line of sight.

Assuming that the $\gamma$-rays and X--rays from \es\ are produced in the 
same region, it is possible to calculate the opacity for pair production 
$\tau_{\gamma\gamma}$ from the source sizes inferred from the 
USA and RXTE/PCA data. The formula given by equation (3) in \citet{mat} 
for the optical depth for an outflow which is nonrelativistic in 
its comoving frame, as corrected by \citet{madej}, is
$$
\tau~=~ 2\times10^3(1+z)^{2\alpha}(1+z-\sqrt{1+z})^2h^{-2}_{60} T_5^{-1}
\times \frac{F_{\mathrm{keV}}}{\mu\textrm{Jy}} \big(
\frac{E_\gamma}{\textrm{GeV}}\big)^{\alpha}
$$
where $T_5$ is the doubling time in units of $10^5$ s and $h_{60}$ the
reduced Hubble constant in units of 60 km s$^{-1}$ Mpc$^{-1}$. 

Using the parameters found in \es\ ($z=0.048$, $\alpha = 1.4$, 
$T_5=2.5$ and $F_\mathrm{keV} = 40\mu$Jy), the opacity for 1 GeV 
photons would be $\tau_{\gamma\gamma}\sim 20$, and $\delta> 1.6$ is 
required in order to have $\tau_{\gamma\gamma}(E_\gamma>1\mathrm{GeV})<1$.
The limit on the required Doppler factor is less than that required 
for Mkn 421 ($\delta=5$), which is not surprising since timescales 
of $T=0.5$ days and a slightly higher luminosity are involved in that object. 

Strictly speaking, the above {$\tau_{\gamma\gamma}$} argument for
anisotropy only applies if the $\gamma$-ray emitting zone is the same
as the soft X--ray-emitting region and, for now, we have no
clear observational indications that this is the case. This is
important, since the jets are likely to be inhomogeneous. However,
since $\gamma$-ray variability on timescales shorter than a day has been
measured -- and since there is an obvious correlation between
X--rays and $\gamma$-rays that has been seen in objects such as Mkn
421 and Mkn 501 -- for such sources {$\delta>$} a few can be deduced as
well.  Relativistic source motion, however, does not avoid the problem of 
the gamma-ray pair production on the external, unbeamed photons likely to 
be present in the environments currently envisioned for the
central engines of such sources.  

The X--ray data presented above imply that the X--ray spectrum 
of \es\ hardens as the source brightens.  This is often measured in 
BL Lac objects;  a hardening of the spectrum when flares
occur, and a blueward shift of the peak of the synchrotron emission 
(and presumably higher energy inverse-Compton emission) by factors 
that can be as large as 100 was measured in the cases of 
Mkn 501 \citep{pian}, 1ES 1426$+$428 and PKS 0548$-$322 \citep{costa}.
In the case of PKS 2005-489 \citep{perl}, a more moderate shift of a
factor of 3 or less of the synchrotron emission was found.  

\subsection{Synchrotron Models and Inferred Parameters}

The spectral change is best illustrated as a correlation between flux
and the photon index. Using PCA data this correlation is illustrated in Figure \ref{plotfour}. 
Even though the two observations were separated by a month and are
certainly related to two different flares,  
it is still interesting to compare this spectral evolution since the 
time series have similar rise and fall timescales, which are also 
comparable to what is seen in the USA detector. It is thus likely  
that the two flares originate from a similar mechanism and that 
the correlation plot has some validity.  The ``clockwise 
motion'' (shown with arrows) observed in the data for \es\, 
has also been seen in flares in Mkn 421
\citep{takah,taka}, in PKS 2155$-$304 \citep{semb} 
and in H0323$+$022 \citep{kohm}, although in
some cases counterclockwise patterns have also been seen (Mkn 501, \citet{catane}). The spectrum 
steepens more rapidly than the flux in the declining phase and hardens 
rapidly in the brightening phase, indicating that the variations of 
the hard X--rays occur faster than those in the soft X--rays both during the
increase and the decrease of the brightness of the source. The
spectral index change of $14\%$ seen here is comparable to the $10\%$
seen in Mkn 421.  The variation observed in the flux-index plane can
provide information about the acceleration process \citep{kirk}. 
Counter-clockwise patterns are expected when acceleration, variability 
and cooling timescales are similar in a flare.  In this case the acceleration
process proceeds from low energy to high energy changing the number of 
particles and making the softer energies vary first.  Clockwise
patterns, where the harder energies vary first, can be explained in
flares where the variability and acceleration timescales are much
less than the cooling timescale.  

For a homogeneous emitting region, the radiative lifetime of a
relativistic electron emitting synchrotron photons with energy
{$E_\mathrm{keV}$} is (in the observer's frame) {$\tau_\mathrm{sync} =
1.2\times 10^3 B^{-3/2}E_\mathrm{keV}^{-1/2}\delta^{-1/2}$} s
\citep{ryb}.  This should give some estimate of the magnetic field $B$,
even though the extent to which the timescale of the flux decrease was 
due to the propagation of the signal throughout the source and to what 
extent it was caused by the synchrotron cooling is not known.  However, 
it is possible to measure the relative decrease of the flux 
{$\Delta F/F$} in three energy bands in a time {$\Delta T$} 
using the PCA data as seen in Figure \ref{plotthree}.  To estimate 
the timescale for a drop by a factor of two in each energy band, 
the measured timescale is divided by a factor {$2\Delta F/F$}.  
The factor {$\Delta F/F$} is smallest for the lowest energy band (24\%) as
expected. It is now possible to find {$\tau_{1/2}(E) =$} 7.7, 4.9 and
$1.9\times 10^5$ s, respectively, for 3, 7 and 12 keV
photons.  According to \citet{taka} we write 
{$\tau_\mathrm{sync}(E) - \tau_\mathrm{sync}(12 \mathrm{keV}) =
1.2\times10^3B^{-3/2}\delta^{-1/2}(E^{-1/2}_\mathrm{keV} -
12^{-1/2})$}.  Comparing the decline of the flux of 3 keV and 
12 keV photons yield {$B=0.007\delta^{-1/3}$}G, while the decay of
the 7 keV and 12 keV photons yield a similar result of
{$B=0.005\delta^{-1/3}$}G.  Using {$\delta = 1.6$} we infer
$B=4\times10^{-3}$ G. It is interesting to note here that this is a
similar value to that found by \citet{perl} (P99 hereafter) in an outburst of PKS
2005$-$489 where variability on timescales of days was observed. 

The peak observed frequency of the synchrotron emission $\nu_s$ of an
electron with {$\gamma_\mathrm{el}$} is given by {$\nu_s \simeq
1.2\times10^6B\gamma_\mathrm{el} ^2 \delta^{-1} $} Hz. Using the
magnetic field inferred above, the Lorentz factors of the electrons
{$\gamma_\mathrm{el}$} radiating at energy $E$ can be estimated from
{$E=2\times10^{-14}\gamma_\mathrm{el} ^2\delta$} keV. This implies that
{$\gamma_\mathrm{el}$} of electrons 
\footnote{Note that {$\gamma_\mathrm{el}$} refers to the 
Lorentz factors of individual radiating particles as distinct from 
the jet bulk Lorentz factor $\Gamma$} radiating in the X--ray band is {$\sim$ $10^7$}.

The value of $B$ calculated as above is significantly lower than 0.2
G, and {$\gamma_\mathrm{el}$} higher than $5\times10^5$ inferred for Mkn 
421 by \citet{taka}, but the values inferred by us are similar to those 
quoted for PKS 2005$-$489 by P99 where a possible 
undersampling was invoked to explain the possibility of faster (shorter 
than $\sim$ 1 day) variability, which in turn would make the inferred 
value for $B$ a lower limit and {$\gamma_\mathrm{el}$} an upper limit. 
We believe that in \es\, the flux is sampled relatively densely, and 
thus the variability is unlikely to be undersampled.  This means that 
the difference in the inferred values of $B$ and {$\gamma_\mathrm{el}$} 
in \es\ as compared to Mkn 421 is unlikely to be a result of the source being 
more compact, while this could be the case for the potentially undersampled 
data for PKS 2005$-$489 (P99).  With our inferred source parameters 
being so close to the values found for PKS 2005$-$489, it is now quite 
possible that real differences between these two sources and Mkn 421 exist.  
A conclusive test of the physical parameters in this source would be 
a clear detection of the TeV $\gamma$-ray emission, and any correlation 
with the X--ray flux.  Also, longer, more sensitive and well-sampled 
observations in the X--ray band are needed to either confirm or disprove 
that the variability pattern of \es\ and its spectral properties 
are as described here. A noticeable difference between PKS 2005$-$489
and \es\ 
is that whereas in the latter a variation in the spectral index {$\Delta \alpha = 0.6$} 
was associated with a flux change of a factor of 30, in the former 
a $\Delta \alpha = 0.35$ produced only a 65\% change in the flux.  
Also note in the USA data, a doubling of the hardness ratio is 
accompanied by a factor of 6 increase in flux. 

\subsection{Continuous Emission: Knot Radiation?} 

The USA lightcurve exhibits a non-zero X--ray flux outside of
the flaring events of a few 10$^{-11}$ erg cm$^{-2}$ s$^{-1}$ (or a
few mCrab, taking 1 Crab $\approx$ 1.7$\times$10$^{-8}$ erg cm$^{-2}$
s$^{-1}$ in the 2--10 keV band). The existence of a steady underlying emission in 
at least one other BL Lac object, Mkn 421, has been invoked in order
to obtain meaningful fits with an exponential decay to X--ray  
flares \citep{fos}.  It is intriguing to investigate if such steady 
flux could originate in more extended jets such as those recently 
resolved by the \textit{Chandra} and \textit{XMM} telescopes, but 
aligned more closely to our line of sight and thus brighter.  
Such knots (or hotspots) in large-scale (hundreds of 
parsecs or more) jets have indeed been seen from radio 
to X--ray energies in many \textit{non-aligned} sources i.e. 
sources where the jet is sufficiently misaligned to allow us to resolve 
the structure of the jet;  of course such structures must also 
originate on a relatively large spatial scale as compared to the 
sub-parsec jets responsible for the rapid, day-scale variability.  
These knots are persistent structures visible on timescales of years, 
and individual spectral energy distributions (SEDs) have been 
established for knots in multiple sources such as M87 \citep{marsh} 
or PKS 0637$-$752 \citep{char}.  Their fluxes are usually a fraction 
of the flux arising from the unresolved core, but a closer alignment 
to our line of sight than for those sources resolved by Chandra and XMM would 
result in greater Doppler boost and could provide a continuous background 
seen in BL Lac-type objects, presumably the objects most closely aligned 
to our line of sight.  

To verify if a flux of a few 10$^{-11}$ erg cm$^{-2}$ s$^{-1}$ could
originate from aligned X--ray knots, some fluxes are taken from the
literature for the cases where the angles to the line of sight ($\theta$ hereafter) and Lorentz 
factors are available from radio observations.  With this, it is possible to
estimate the flux enhancement for $\theta \sim 0^\circ$ and the 
luminosity at a distance similar to that of \es. The observed flux
would depend on three factors: \begin{itemize}
\item{A change in Doppler boost that multiplies the flux $F$ by
{$(\tilde{\delta}/\delta)^4$} (see \citet{ursha} for luminosity
conversions) where {$\tilde{\delta}$} is the Doppler
factor for the same source but with an angle {$\theta =
0^\circ$}. Thus the amplification would be {\large$\big[\frac{1-\beta \cos \theta}{1-\beta}\big]^4$}. Knowing the Lorentz factor $\Gamma$ we can write
$\beta^2$={\large$\frac{\Gamma^2 -1}{\Gamma^2}$} }
\item{The distance difference changes the flux by a factor of
{$(z/0.048)^2$}}
\item{ a K-correction has to be applied since the bandwidth is narrowed,
changing the flux by a factor of
{\large$\big(\frac{1+z}{1.048}\big)^{\alpha-1}$} where $\alpha$ is the spectral index. This effect
is minimal compared to the previous ones since in the X--ray band
usually $\alpha \geq 1$ for BL Lac spectra. } 
\end{itemize} 

The final ratio $A$ of the observed fluxes is 
then the product of these three terms.

\subsubsection{Comparison with M87}
The most popular scenario for the ``parent population'' 
of XBL-type blazars such as \es\ is that these objects are FR-I type
radio galaxies  
with their jets aligned close to our line of sight (see, e.g., 
Urry, Padovani, \& Stickel 1991, UPS91 hereafter).  In contrast, PKS
0637-752 is 
a member of the intrinsically more powerful FR-II class of objects.  
We thus believe that it is appropriate to compare \es\ with the 
FR I radio galaxy M87 ($z=0.004$), a nearby object 
that has knots resolved in the X--ray band \citep{marsh} as
well as in the radio and optical (e.g. \citet{perlmana} and references
therein). In the most popular models the X--ray radiation is either synchrotron
radiation or inverse Compton scattering (see e.g. \citet{yang}).  From \citet{bir} (B99 hereafter) we find that a relativistic jet model requires a bulk
flow with Lorentz factor $\Gamma \geq 6$ and a jet orientation within
$\theta \leq 19^\circ$ from the line of sight. We will use two sets of
$(\theta,\Gamma)$ presented in B99 based on this model.
 
The set $(\theta,\Gamma)\sim(18^\circ,12)$ is a possible configuration 
of the jet. These values may offer an explanation for the 
apparent lack of superluminal motion in M87 on parsec scales, and 
were assumed in the model of \citet{bick}. From \citet{marsh} the 
flux density at 1 keV, summed over all the knots, is approximately 
380 nJy and $\alpha=1.46$. Using a distance of 16 Mpc for M87, $H_0$ =
60 km s$^{-1}$ Mpc$^{-1}$ and $q_0$=0.1, then $A \sim 250$ 
would yield $\sim$ 20$\times$ 10$^{-11}$ erg cm$^{-2}$ s$^{-1}$ (or 8
mCrab) of flux coming from the knots.   
In this case, using an angle of $\sim4^\circ$ would reduce the 
flux ratio by a factor of 8 (from 250) which would still keep the boosted 
knot radiation in the detectable range.    
An angle of $4^\circ$ was used as well in this example since this
is the mean value of angle to the line of sight used recently by
\citet{costaghi} as an input parameter for the successful 
synchrotron self-compton (SSC) model those authors have used to 
predict TeV candidate BL Lac objects.

The set of values $(\theta,\Gamma)\sim(16^\circ,8)$ also presented in
B99 and consistent with their results yield a flux
coming from the knots that is 40 times smaller (or 0.2 mCrab). This
configuration is thus less likely to result in a flux of the level
that is being looked for.

\subsubsection{Comparison with 3C66B}

3C66B is a low-luminosity FRI radio galaxy ($z=0.0215$). Its jet has knots
resolved in the radio, optical and X--ray band (\citet{hardcastle} and
references therein). From \citet{giov} we find that $\theta$ seems
well constrained (about $45^\circ$) but that $\Gamma$ is not (between
1 and 7).

The set $(\theta,\Gamma)\sim(45^\circ,7)$ is a possible configuration of the
jet that would come within the right order of magnitude to explain the
origin of the X--ray radiation by IC scattering of seed photons from a
hidden BL Lac in the nucleus of 3C66B. From \citet{hardcastle} the
flux density at 1 keV, for the 2 brightest knots out of 5 (A and B),
is approximately 10 nJy. A value of $\alpha=1.31$ is used, the one
found for the jet, though the exact value does not matter much given
the comparable redshift of 3C66B to the one of \es. The inferred value
of $A \sim 1.6 \times 10^5$ yields $\sim$ 4$\times$ 10$^{-9}$ erg
cm$^{-2}$ s$^{-1}$ (or 230 mCrab) of flux coming from the knots in the
case of an angle at $0^\circ$. Using an angle of $\sim4^\circ$ would
reduce the flux ratio by a factor of 2 only and make it still a bright steady
source. The level of expected radiation in this case is actually so
high that a similar object pointing closer to the line of sight would
not be unnoticed in radio emission, thus the
$(\theta,\Gamma)\sim(45^\circ,7)$ set seems quite extreme.   

Taking a slower jet ($\beta \sim 0.75$) also mentioned in
\citet{hardcastle} the flux coming from the knots of 3C66B would be a
factor of $6\times10^4$ lower than in the previous case, significantly
below the level of a few 10$^{-11}$ erg cm$^{-2}$ s$^{-1}$.

The results from M87 and 3C66B show that it is possible that radiation from knots 
moving close to the line of sight can generate a significant fraction of 
the baseline level of X--ray flux invoked in at least two cases, Mkn 421 
and \es\ here, assuming they contain knots similar to those observed in 
M87 or 3C66B. Recent VLBI
observations show that the line of sight angle with Mkn 421 is
constrained to be in the $0^\circ - 30^\circ$ range \citep{giov}, thus
a small angle to the line of sight is a possibility for this source. 
In a more general way,
UPS91 predict that FR I radio galaxies should have jets with bulk 
flow speeds in the range from $\Gamma \sim$ 5 to $\sim$ 35, with most
near $\Gamma \sim$ 7, and they derive a critical angle
$\theta_\textrm{crit} \sim 10^\circ$ for the FR I/BL Lac division. If
bright knots are common in BL Lac objects then the amplification
factors associated with the large Lorentz factors invoked here could
generate a continuous background at the level mentioned above
(assuming $\theta$ is of the order of a few degrees only). A more
thorough 
search for evidence of continuous emission in blazars and other
signatures of boosted knot radiation is currently under way.    

\section{Conclusions} 
This paper reports X--ray data obtained with the USA and RXTE missions, 
as well as archival observations for the BL Lac object \es.
Variability on the timescale of a few days with a threefold flux 
increase in the 1--16 keV band was observed with the USA detector;  this is 
consistent with timescales that have been seen in the optical band 
at other times.  The X--ray data also show a clear correlation of the 
X--ray flux with the spectrum, which becomes harder when the source 
is brighter.  The data presented in this paper represent an important 
contribution to the study of this object, a potential TeV emitter, 
as long term monitoring observations of it in the X--ray band 
were seldom performed in the past. 

From the X--ray variability time scales, we estimate the Doppler
factor of the jet to be $>$ 1.6.  The spectral variability data 
allow an estimate of the magnetic field which is lower ($\sim$ 
milliGauss), and Lorentz factor of radiating electrons which is 
higher ($\sim 10^7$) than those derived for the prototype TeV blazar 
Mkn 421.  These parameters are more in line with those inferred 
for another possible TeV - emitting blazar, PKS 2005$-$489.  The 
data also show the need for X--ray observations of \es\ longer 
than a few days in order to more accurately characterize the timescale 
of variability and its relationship with the spectrum. Also, since the source
has been seen to be extremely variable in the optical band, future 
multi-wavelength observations of this object should include
simultaneous optical observations. 

We attempted to give a plausible scenario explaining the apparent baseline
(``quiescent'') emission level seen here in \es\ but also possibly
previously observed in Mkn 421. Jets containing bright knots such as
those seen by \textit{Chandra}, but aligned more closely, within a few degrees to the line of sight (and therefore 
not resolvable as easily as jets with larger angles) and with Lorentz
factors of $\sim 10$ have the required  
boosted flux to account for the ``quiescent'' X--ray emission in 
XBL-type BL Lac objects.  Such emission would persist on timescales 
much longer than the duration of the flares that are much more likely 
to occur closer to the central object.  This scenario can be confirmed 
by a more systematic study of knot X--ray luminosities in FR I
objects, as well as the currently ongoing population 
study of the steady component of X--ray emission in BL Lac objects.     

\acknowledgements 

We are grateful to Paul Kunz for providing useful analysis
software. Our referee, Eric Perlman, contributed useful comments that
resulted in broadening the scope and clarifying some points in the
paper, for which we thank him. This work was performed while RMB held a National Research
Council Research Associateship Award at NRL. JDS is grateful to the NASA  
Applied Information Technology Research Program for support. Work at
SLAC was supported by the Department of Energy contract 
to Stanford University DE-AC03-76SF00515. Basic research in X--ray 
Astronomy at the Naval Research Laboratory is supported by
ONR/NRL. This paper made use of quick-look results provided by the ASM/RXTE 
team (see \url{http://xte.mit.edu}), as well as data provided by HEASARC, a 
service of NASA/Goddard Space Flight Center.

\begin{deluxetable}{ccc}
\tablewidth{0pc}
\tablecolumns{4}
\tablecaption{Fluxes and Hardness Ratios in the USA Experiment Data for BL Lac Object \es}
\tablehead{
\colhead{Date (2000)} & \colhead{$F$ (mCrab)}  & \colhead{HR} \\
&  &  
}

\startdata

Oct 6&  4.2&    $0.70\pm 0.03$  \\
Oct 6&  4.1&    $0.70\pm 0.03$  \\
Oct 8&  5.2&    $0.76 \pm 0.03$ \\
Oct 8&  5.2&    $0.71 \pm 0.02$ \\
Oct 9&  5.0&    $0.61 \pm 0.02$ \\
Oct 10& 5.3&    $0.63 \pm 0.03$ \\
Oct 11&         4.0&    $0.59 \pm 0.03$ \\
Oct 12&         3.5&    $0.69 \pm 0.04$ \\
Oct 14&  2.7&   $0.48 \pm 0.04$ \\
Oct 15&  3.2&   $0.69 \pm 0.03$ \\
Oct 16&  3.9&   $0.71 \pm 0.03$ \\
Oct 17&  3.9&   $0.63 \pm 0.04$ \\
Oct 19&  4.1&   $0.76 \pm 0.04$ \\
Oct 20&  3.3&   $0.72 \pm 0.04$ \\
Oct 22&  2.6&   $0.49 \pm 0.04$ \\
Oct 23&  2.5&   $0.55 \pm 0.04$ \\
Oct 24&  2.7&   $0.60 \pm 0.03$ \\
Oct 27&  4.1&   $0.66 \pm 0.03$ \\
Oct 28&  4.1&   $0.70 \pm 0.03$ \\
Oct 29&  4.7&    $0.79 \pm 0.03$ \\
Nov 2&   3.7&    $0.62 \pm 0.04$ \\
Nov 3&   3.4&    $0.58 \pm 0.03$ \\
Nov 5&   3.9&    $0.66 \pm 0.03$ \\
Nov 7&   4.4&    $0.63 \pm 0.04$ \\
Nov 8&   3.9&    $0.68 \pm 0.03$ \\
Nov 9&   4.9&    $0.71 \pm 0.04$ \\
Nov 10&  6.0&    $0.80 \pm 0.05$ \\
Nov 11&  8.7&    $0.79 \pm 0.03$ \\
Nov 12&  10.4&   $0.90 \pm 0.03$ \\
Nov 14&  11.6&   $0.99 \pm 0.03$ \\ 
Nov 15&  10.8&   $0.90 \pm 0.03$ \\

\enddata

\end{deluxetable}

\begin{deluxetable}{ccc}
\tablewidth{0pc}
\tablecolumns{4}
\tablecaption{Single Power Law Spectral Fits to the RXTE PCA data}
\tablehead{
\colhead{Date (2000)} & \colhead{$F$ (2-10 keV)}  & 
\colhead{$\alpha$} \\
 & \colhead {$(\times 10^{-11} {\rm ~erg ~cm^{-2} ~s^{-1}})$} 
& \colhead {$(F_\nu \propto \nu^{-\alpha})$}
}

\startdata

Jul 28.9&       1.40&   $1.42\pm 0.05$  \\
Jul 29.0&       1.38&   $1.38\pm 0.05$  \\
Jul 29.1&       1.40&   $1.39 \pm 0.05$ \\
Jul 29.9&       1.32&   $1.57 \pm 0.06$ \\
Jul 30.0&       1.28&   $1.45 \pm 0.06$ \\
Jul 30.1&       1.27&   $1.51 \pm 0.06$ \\
Jul 30.9&       1.19&   $1.65 \pm 0.06$ \\
Aug 01.0&       1.22&   $1.63 \pm 0.06$ \\
Aug 01.1&   1.19&   $1.6  \pm 0.06$ \\
Aug 01.9&   1.06&   $1.66 \pm 0.07$ \\
Aug 02.0&   1.02&   $1.66 \pm 0.07$ \\
Aug 02.1&   1.08&   $1.67 \pm 0.07$ \\
Sept 01.0&   0.86&   $1.68 \pm 0.04$ \\
Sept 02.9&   0.84&   $1.62 \pm 0.04$ \\
Sept 03.9&   1.04&   $1.53 \pm 0.04$ \\
Sept 04.9&   1.19&   $1.51 \pm 0.04$ \\
Sept 05.9&   1.38&   $1.37 \pm 0.04$ \\

\enddata

\end{deluxetable}

\clearpage

\end{document}